\documentclass[superscriptaddress, amsmath,amssymb, pre, twocolumn,floatfix]{revtex4-1}

\usepackage{graphicx}
\usepackage{dcolumn}
\usepackage{bm}
\usepackage{hyperref}
\usepackage{marginnote}
\usepackage{subfigure}
\newcommand{\Z}{\langle z \rangle}
\usepackage{color}

\begin{document}
\title{Rigidity percolation control of the brittle-ductile transition in disordered networks}

\author{Estelle Berthier$^*$}
\affiliation{Dept. of Physics, North Carolina State University, Raleigh, NC, USA}
\email{corresponding author: Estelle Berthier (ehberthi@ncsu.edu)}
\author{Jonathan E. Kollmer}
\affiliation{Dept. of Physics, North Carolina State University, Raleigh, NC, USA}
\author{Silke E. Henkes}
\affiliation{School of Mathematics, University of Bristol, Bristol, UK}
\author{Kuang Liu}
\affiliation{Dept. of Physics, Syracuse University, Syracuse, NY, USA}
\author{Jennifer M. Schwarz}
\affiliation{Dept. of Physics, Syracuse University, Syracuse, NY, USA}
\author{Karen E. Daniels}
\affiliation{Dept. of Physics, North Carolina State University, Raleigh, NC, USA}

\begin{abstract}
In ordinary solids, material disorder is known to increase the size of the process zone in which stress concentrates at the crack tip, causing a transition from localized to diffuse failure. Here, we report experiments on disordered 2D lattices, derived from frictional particle packings, in which the mean coordination number $\Z$ of the underlying network provides a similar control. Our experiments show that tuning the connectivity of the network provides access to a range of behaviors from brittle to ductile failure. We elucidate the cooperative origins of this transition using a frictional pebble game algorithm on the original, intact lattices. We find that the transition corresponds to the isostatic value $\Z = 3$ in the large-friction limit, with brittle failure occurring for structures vertically spanned by a rigid cluster, and ductile failure for floppy networks containing nonspanning rigid clusters. Furthermore, we find that  individual failure events typically occur within the floppy regions separated by the rigid clusters. 
\end{abstract}

\maketitle

Materials as varied as semiconductors \cite{oshima_extraordinary_2018}, aqueous foams \cite{arif_spontaneous_2012}, polymers \cite{jang_ductile-brittle_1984}, metals \cite{Tanguy2005} and rocks \cite{wong_brittle-ductile_2012} exhibit a brittle-ductile transition  when parameters such as geometry, temperature, pressure, loading rate, or even illumination are varied. Because brittle failure occurs suddenly and unpredictably, often leading to catastrophic effects, it is important to understand which underlying control parameters are able to tune the failure behavior of materials and structures, including those that are heterogeneous and/or hierarchical. Improvements in the prediction of failure modes are essential to the design of optimal properties.

Controlled studies of the brittle-ductile transition have been achieved both numerically and experimentally. For example, increasing material disorder has been shown  \cite{roux_rupture_1988,alava_role_2008, shekhawat_damage_2013,curtin_brittle_1990,kahng_electrical_1988} to increase the fracture process zone (FPZ) size, resulting in a less concentrated stress at the crick tip. The size of the FPZ eventually diverges for infinite disorder, producing a transition from a brittle narrow crack type failure to percolation-like diffuse behavior. Experimentally, Hanifpour \textit{et al.} \cite{hanifpour_mechanics_2018} showed that 3D-printed disordered auxetic lattices can fail in a ductile or brittle (with disorder-dependent tensile strength) manner depending on the loading direction.  Driscoll \textit{et al.} \cite{driscoll_role_2016}  demonstrated the key role of material rigidity in experiments on weakly-disordered honeycomb acrylic lattices, and also the key role of connectivity in numerical studies of two different types of spring networks, one of which was derived from numerical realizations of frictionless packings. Numerical studies in Zhang \textit{et al.} \cite{zhang_fiber_2017}, on the other hand, focused on how the nonlinear alignment of springs in a randomly diluted triangular lattice controls the transition at connectivities below what is known as the central force rigidity percolation point. Finally, Bouzid and Del Gado \cite{bouzid_network_2017} focused on  the role of connectivity  in a disordered system, by studying the mechanical properties of simulated soft gels, showing that the topology greatly affects stress redistribution and deformation of the branches, resulting in brittle failure of the more homogeneous stiffer gels with higher mean connectivity.

Here, we focus on experimentally-addressing this last question -- tuning connectivity within an inherently disordered experimental system -- by isolating the effects of the mean coordination number $\Z$. We show that this parameter controls access to a variety of behaviors, from diffuse to brittle failure, in experimentally fabricated disordered 2D lattices made of stiff acrylic. This determination is particularly important for a large class of disordered materials: solid lattices or rigid foams composed of disordered beams bonded at their intersections. These materials present an attractive choice for producing lightweight, tunable structures and have recently been developed as (mechanical) metamaterials
\cite{Paulose7639,bertoldi_flexible_2017,goodrich_principle_2015,reid_auxetic_2018,rocks_designing_2017,hanifpour_mechanics_2018}. 
Furthermore, such discrete structures have been commonly used in statistical models of fracture \cite{alava_statistical_2006}, as a means to conveniently discretize a material. For example, the random fuse model (RFM) \cite{de_arcangelis_random_1985,gilabert_random_1987,alava_role_2008}, fiber bundle model (FBM) \cite{pradhan_failure_2010,Kun2006} or elastic springs model \cite{alava_role_2008,curtin_brittle_1990,zhang_fiber_2017,nukala_percolation_2004,ray_breakdown_2006} descriptions rely on such lattices in which material disorder is introduced. Despite a wealth of models, few fracture experiments to discriminate between the various models have been performed on disordered lattices \cite{driscoll_role_2016,hanifpour_mechanics_2018}, in part due to the difficulty of creating samples by hand \cite{gilabert_random_1987}.

\begin{figure*}
\center
\begin{tabular}{ccc}
\subfigure[]{
\includegraphics[width=.3\textwidth]{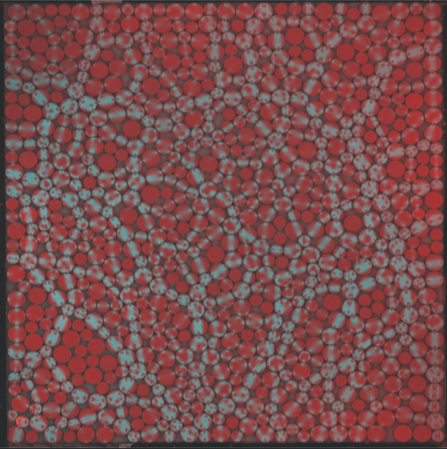} 
}&
\subfigure[]{
\includegraphics[width=.3\textwidth]{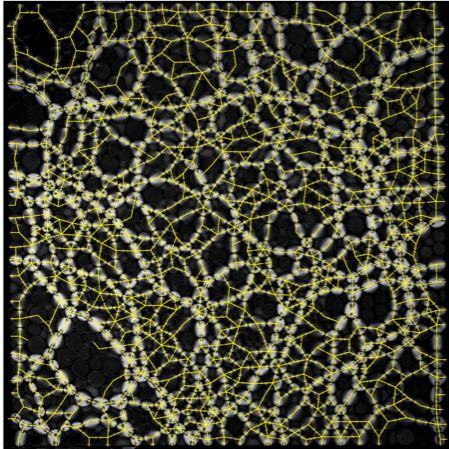} 
}\ &
\subfigure[]{
\includegraphics[width=.3\textwidth]{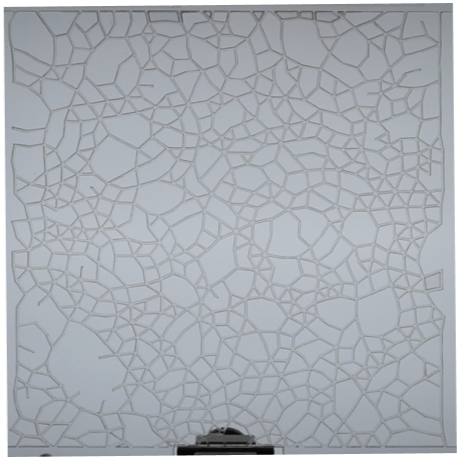} 
}\ 
\end{tabular}
\caption{{\bf Creating a disordered lattice from a frictional granular packing.} (a) Force chains (cyan) recorded through a polariscope image  of a  2D assembly of frictional photoelastic disks of two sizes. (b) Contact network (yellow lines), as extracted using our open source photoelastic solver \cite{jekollmer_pegs:_2018}, and shown on the resulting reconstructed image \cite{daniels_photoelastic_2017}. (c) Corresponding physical structure, laser-cut from an acrylic plate.}
\label{fig:makelattice}
\end{figure*}

With the advent of laser-cutting technology, we are able to readily produce samples with precise, reproducible properties and thereby perform controlled experiments on the failure behavior of disordered lattices with various connectivities. We focus on the limiting case where disorder is primarily set via the geometry of the lattice,  and material fluctuations are intended to be negligible. Rather than diluting \cite{broedersz_criticality_2011,ellenbroek_rigidity_2015,feng_percolation_1984,zhang_fiber_2017,driscoll_role_2016} and perturbing \cite{driscoll_role_2016} a regular lattice (where the underlying crystalline structure is still present), we instead draw on \textit{experimentally-generated}, inherently disordered materials, as shown in Fig.~\ref{fig:makelattice}. This approach is inspired by prior studies on numerically-generated packings of jammed frictionless particles \cite{driscoll_role_2016,goodrich_principle_2015,ellenbroek_non-affine_2009,ellenbroek_rigidity_2015}.
Using such packing-derived networks has the additional benefit of allowing connections to some particular features of jammed packings \cite{liu_jamming_2010,van_hecke_jamming_2010}. For example, while their shear modulus vanishes at the jamming point, their bulk modulus remains finite \cite{ellenbroek_rigidity_2015}, which is rather different from, for example, central-force spring networks where both the bulk and shear modulus vanish at their equivalent jamming point \cite{ellenbroek_non-affine_2009}. 

Our networks are generated from real, frictional packings \cite{majmudar_contact_2005} drawn from experiments on uniaxially compressed quasi-2D bidisperse granular packings of frictional photoelastic disks, created using the techniques described in \cite{puckett_equilibrating_2013,daniels_photoelastic_2017,kollmer_betweenness_2018,bililign2018}. In each case, the test structure has a geometry defined by the binary contact network of these disks. Each particle center is set to be a crosslink (node), and each particle contact in the original packing is a beam (edge).  The beams are of uniform thickness and connect at the crosslinks. This process is shown in Fig.~\ref{fig:makelattice}, resulting in a disordered lattice that can be cut from a single acrylic sheet. Once cut, each sample can be exposed to either tensile or compressive tests through to the point of successive failures.

In our experiments, we measure the spatial and temporal pattern of failure as a function of the connectivity of the disordered lattice. We find a transition between ductile and brittle behaviors as a function of the coordination number $\Z$. This is characterized by both the force response to compressive or tensile deformation, and the magnitude of the resulting force drops. In the brittle samples, failure develops along narrow, meandering cracks until the sample breaks in two.

We use rigidity theory to interpret changes in these patterns. 
It has been successfully applied to disordered networks of linear springs 
\cite{moukarzel_stressed_1995,jacobs_generic_1995} 
 and to granular packings mapped to a node-spring network with 2 or 3 degrees of freedom at each node \cite{ellenbroek_rigidity_2015,ellenbroek_non-affine_2009,feng_percolation_1985,henkes_rigid_2016}. 
To identify rigid clusters based on the topology of the spring network of frictionless packings, one uses a $(2,3)$ pebble game  \cite{jacobs_algorithm_1997}.
However, our disordered 2D lattices originate from frictional packings, which add particle rotations and tangential forces to the central forces, so that we use a  $(3,3)$ pebble game recently developed for the study of \textit{frictional} granular packings \cite{henkes_rigid_2016}.

We associate the central forces to the stretching/compressing of the beams and the tangential ones to the rotations around welded points.  Arguing that these deformations dominate the response in our networks, we find that the frictional rigid cluster decomposition allows us to differentiate the response of low- vs. high-connectivity networks and capture the change in failure behavior from ductile to brittle.  We are also able to establish that in the presence of both rigid and floppy clusters for networks near the transition point, identified close to 
$\Z_{\mathrm{iso}} = 3$,
which is the isostatic connectivity of an infinite friction coefficient packing,  failures primarily occur on floppy edges located between the rigid clusters.

\section*{Methods \label{sec:methods}}

Each sample's geometry is derived from a force chain network observed in an experimental 2D granular material, as shown in Fig.~\ref{fig:makelattice}(a). The apparatus used to obtain these configurations contains $N=824$ to $890$ bidisperse particles (two distinct radii, 5.5 mm and 7.7 mm, in approximately equal numbers) made from photoelastic Vishay PhotoStress circular disks. The mixture is placed on an air-table to remove basal friction and a cyclic uniaxial loading introduces local contact rearrangements. This allows for the observation of multiple force distributions. We use an open source photoelastic solver  \cite{jekollmer_pegs:_2018} using the method described in \cite{daniels_photoelastic_2017} to obtain the contact network underlying the observed force chains. An example of the binary network obtained by this method is shown in Fig.~\ref{fig:makelattice}(b).

We characterize each network by its mean coordination number $\Z$, which is the mean  number of beams intersecting at the crosslinks in each sample. 
In performing our experiments, we use 8 networks, drawn from 4 different initial granular configurations loaded by different amounts. Two of these networks are obtained using a packing  derived from previously-collected data \cite{bililign2018}.
 Due to the lack of experimental data near that connectivity value, the network with $\Z\approx 3$ was obtained by pruning a higher connectivity network: contacts with a force below a threshold were discarded, the threshold being set so as to obtain the targeted connectivity. Together, this provides  values ranging from $\Z = 2.4$ to $\Z = 3.6$, selected to be approximately evenly spaced in the vicinity of 
$\Z_\mathrm{iso} = 3$. 

To render these networks as physical samples, we laser cut acrylic plastic plates of thickness $h = 3.17$~mm, and approximately $30\times30$~cm$^2$. In the sample, each contact force of the granular system corresponds to a beam of width $1.4$~mm. The beams intersect at crosslinks (particle centers), and are of variable length depending on the radii of the two contacting particles. Thus, the sample has beams with width:length ratios of $0.15$ to $0.2$;  shorter beams are slightly stiffer than the longer ones, for constant elastic modulus $E\approx 3$~GPa.  In addition, some material disorder remains intrinsic to the bulk material, and some defects are presumably created during the laser-cutting process.

\begin{figure}
\center
 \includegraphics[width=0.99\columnwidth]{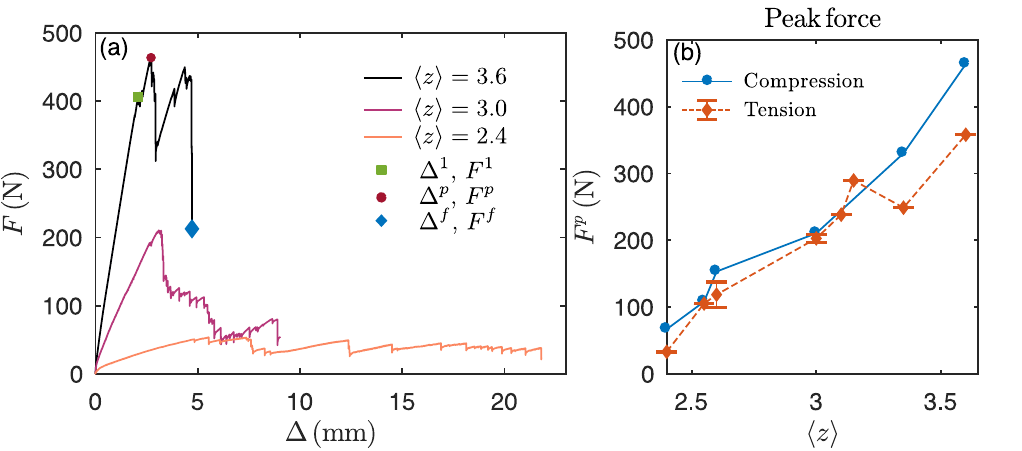} 
\caption{{\bf Bulk response of the disordered lattices.} (a) Effect of $\Z$ on the macroscopic force-displacement response of networks tested in compression. The loading at first failure ($\Delta^1$, $F^1$) is indicated by a green square, the peak loading ($\Delta^p$, $F^p$) by a red dot, and the loading at failure ($\Delta^f$, $F^f$) by a blue diamond. (b) Measured values of $F^p$ as a function of $\Z$, for compression (blue, solid) and tension (red, dashed) experiments.
}
\label{fig:macroResp}
\end{figure}

We test each sample in displacement-controlled experiments using an Instron 5940 Single Column system. Compression experiments are conducted with a displacement rate of $1.5$~mm/min, and tensile experiments at $1.0$~mm/min; both loading rates allow for well separated failure events or avalanches. In compression, the structure is constrained between two transparent acrylic plates to prevent out-of-plane buckling.  To analyze the deformation and failure behaviors, we measure the macroscopic force $F$ as a function of the imposed displacement $\Delta$ and record each experiment using a Nikon D850 digital camera, at a frame rate ranging from 24 to 60 fps. During the course of each experiment, fracture events occur within the sample and sequentially deteriorate its structure through the sudden breaking of one or more beams. Each event appears as a drop in the macroscopic response, as shown in Fig.~\ref{fig:macroResp}(a). The corresponding failure locations are determined from examining the image taken at the corresponding time.

\section*{Results}
\label{sec:results}

\subsection*{Macroscopic behavior}

Several typical macroscopic force-displacement responses measured are shown in Fig.~\ref{fig:macroResp}(a), for three different samples tested in compression. For low applied displacement $\Delta$, there is an initial elastic regime in which deformations are principally accommodated by beam-bending at the nodes. At a displacement $\Delta^1$ and corresponding force $F^1$, the first beam breaks, as shown by the green square in  Fig.~\ref{fig:macroResp}(a) on the curve taken from the $\Z = 3.6$ sample. Each such event  triggers an abrupt force drop $\delta F$. After one or a few such failure events (typically small, i.e. involving one beam breakage), the system reaches a peak force $F^p$, indicated by the red dot in Fig.~\ref{fig:macroResp}(a). From this point onward, the response intermittently switches between two behaviors: elastic deformation, and dissipative failure events. This intermittency continues until a critical loading $\Delta^f$ is reached, marked by a blue diamond, at which the failures have laterally spanned the entire system, such that there is no set of beams connecting the top and bottom boundaries. We end all tests at this point. 

This general behavior is observed for all samples, but the specifics are $\Z$-dependent. As shown in Fig.~\ref{fig:macroResp}(a), networks with higher coordination number are stiffer, as previously observed in simulations \cite{ellenbroek_non-affine_2009,ellenbroek_rigidity_2015}. The peak force  $F^p$ increases with $\Z$, as plotted in Fig.~\ref{fig:macroResp}(b) for both compression and tension experiments. Finally, lower-$\Z$ networks exhibit a different type of intermittency: many small events are observable before failure, rather than a few large events, and some of the force drops correspond to sudden sliding of beams put in contact due to the high compression.

\begin{figure*}
 \includegraphics[width=\textwidth]{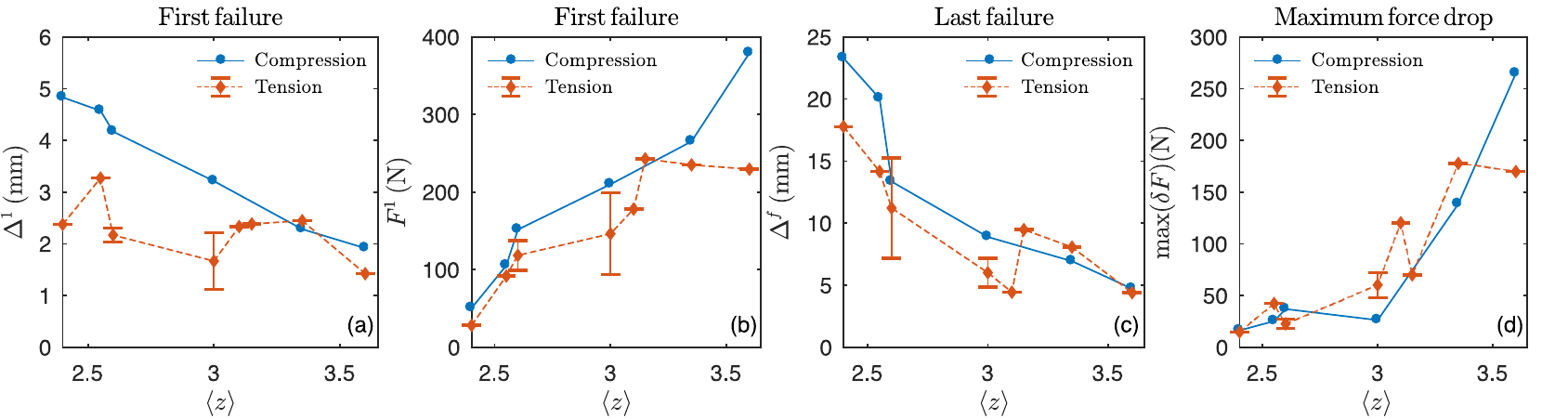} 
\caption{{\bf Characterization of failure events.} Effect of $\Z$ on   (a) the displacement at first failure event, (b) the force at first failure event,  (c) the applied displacement at complete failure of the network, and (d) the maximum force drop occurring during the experiment. Compression (plain blue) and tension (dashed red) results are shown for each sample.}
\label{fig:failEvents}
\end{figure*}

To characterize the failure behavior we consider 4 quantities: the loading $\Delta^1$, its corresponding force $F^1$, the loading at failure $\Delta^f$, and the maximum observed force drop $\mathrm{max}(\delta F)$. Each is shown in Fig.~\ref{fig:failEvents} as a function of $\Z$, for both compression (solid blue lines) and tension (dashed red lines). In compression, we observe that as  $\Z$ increases, the applied displacement $\Delta^1$ decreases, i.e. the first failure event occurs at a lower displacement, while the force at that events rises. The displacement $\Delta^f$ at which there is system-spanning failure is also significantly lower for larger $\Z$.  Finally, for  $\Z\leq 3$, there is progressive damaging of the samples, with typically one or two beams involved in a failure avalanche. For $\Z>3$, there are also large (involving more than 3 beams breakage), catastrophic events, associated with large drops of force. This is captured by the evolution of the maximum force drop $\mathrm{max}(\delta F)$ with $\Z$.

We recall that the network with  $\Z=3.0$ was artificially obtained by removing low-force contacts from the initial, higher-connectivity network. Yet, no peculiar behavior is noticeable from the results: the samples follow the same trends  as the others.
In tension, similar behaviors are obtained, but the data is noisier as a function of $\Z$ and the samples are less resilient than  in compression: premature failure occurs more frequently.

\subsection*{Spatial patterns of failure}

\begin{figure}
\centering
\includegraphics[width=0.85\columnwidth]{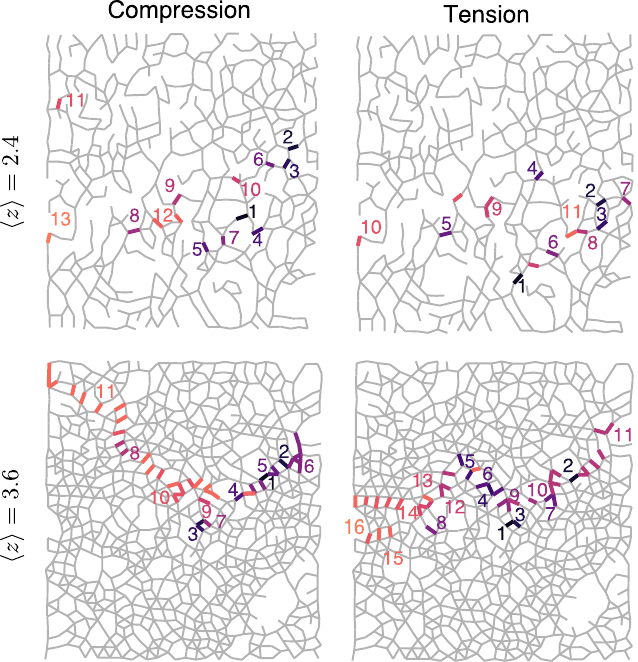} 
\caption{{\bf Spatial patterns of failure.} Thick bars mark each failed beam within the samples, drawn on the intact disordered lattice. The temporal order of each avalanche is numbered  and color-matched to the  beams failing during each failure event. Samples are of average coordination number $\Z = 2.4$ (top) and $\Z = 3.6$ (bottom), for both compression (left) and tension (right).}
\label{fig:FailPatterns}
\end{figure}

We find that the spatial pattern of the individual failed beams varies significantly with $\Z$, as illustrated in Fig.~\ref{fig:FailPatterns} for the two extreme values of $\Z = 2.4$ and $\Z = 3.6$, under both compression and tension. The initial disordered lattice is shown in gray, with each beam that failed during the entire experiment colored as a thick bar. 
For low-connectivity lattices ($\Z<3$), failures are rather spread throughout the sample, while for highly connected networks ($\Z \geq 3$) the failed edges are localized, forming a narrow meandering (especially in tension) crack. 
While this general trend is not affected by the type of loading, we nonetheless observe that the failure locations are not identical in the two cases. From repeating the same experiment with ``identical''  samples, we observe that this can be partly attributed to the sample-to-sample variation in the material disorder, both inherent and introduced during cutting.

\begin{figure}
\centering
\includegraphics[width=0.99\columnwidth]{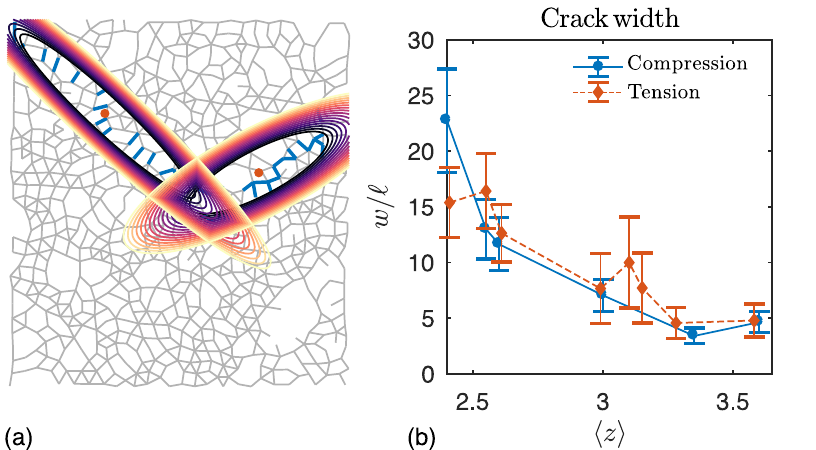} 
\caption{{\bf Characterization of the crack width.} (a) Example of the calculation of crack width $w$ for two ellipses covering  $70\%$ (dark) to $100\%$ (light) of the broken edges. Ellipse centers are marked with an orange dot. 
(b) The dependence of the crack width $w$ as a function of $\Z$. Values are normalized by the minimal edge length $\ell$ and are shown for  compression (solid blue) and tension (dashed red) experiments. }
\label{fig:ellipses}
\end{figure}
 
To characterize these spatial patterns, we define a crack width $w$ as follow: we compute the eigenvalues of the covariance matrix of the cloud of broken points, which we then scale such that the corresponding ellipse, centered on the centroid of the broken edges, encloses $70\%$ (darker ellipses in Fig.~\ref{fig:ellipses}(a))  to $100\%$ (lighter ellipses) of the broken edges. The crack width is defined as the mean value of the scaled smallest eigenvalues (minor axis of the ellipses). In the case of a meandering crack, we use two or more ellipses to cover the straight pieces (see Fig.~\ref{fig:ellipses}a) and take the mean as the crack width. When no particular orientation is noticeable, as in the top row of Fig.~\ref{fig:FailPatterns}, one ellipse is used. 

As suggested by the images in Fig.~\ref{fig:FailPatterns}, $w$, normalized by the smallest beam dimension $\ell$, decreases as a function of $\Z$ (see Fig.~\ref{fig:ellipses}(b)). 
A quantitatively similar trend is observed for both compression and tension experiments.

Therefore, we  identify two types of behavior depending on the connectivity of the lattices. For samples with $\Z < 3$, failure events progress continuously in the system, with a rather broad spatial distribution of failure events. We refer to this behavior as ductile.  As $\Z$ increases, the characteristic parameters (loading at first and last failure event, maximum force drop, crack width) evolve continuously. For $\Z > 3$, a combination of small and large localized catastrophic failure events contribute to the failure of the system via the spanning of a narrow crack. For simplicity, we will refer to this more brittle behavior as ``brittle'', to contrast it with the more ductile behavior observed for lower connectivity.

\section*{Frictional pebble game}

\begin{figure*}
\centering
\includegraphics[width=0.65\textwidth]{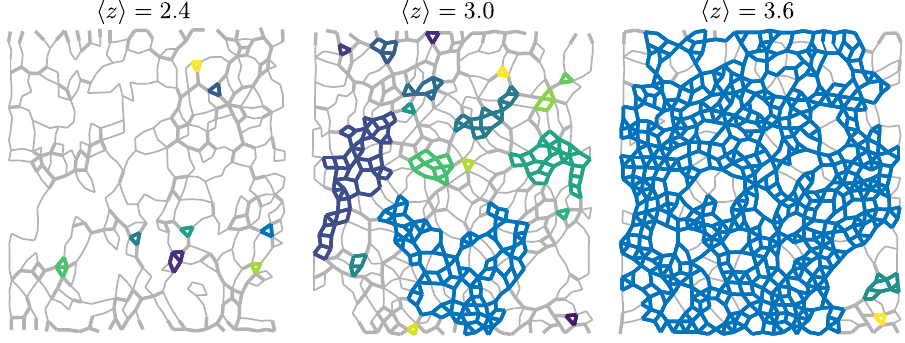}
\caption{{\bf Rigid cluster identification.} Rigid cluster decompositions of three networks with different mean connectivity. The floppy edges are in gray and each color indicates a different rigid cluster. The thicker (thinner) edges designate double (single) bonds in the constraint network.}
\label{fig:clusters}
\end{figure*}

To better understand how this transition emerges, we apply rigidity theory to a mathematical representation inspired by the granular origins of our samples. In creating the lattices, we start from frictional particle packings and essentially ``glue'' the disks together, so that we have effectively increased the friction coefficient to $\mu=\infty$. We have chosen our pebble game parameters to take these permanent attachments into account.

The mean coordination at which the rigidity transition occurs can be estimated by Maxwell constraint counting \cite{maxwell_l._1864} for frictionless granular packings and using a generalized isostaticity criterion introduced by Shundyak \textit{et al.} \cite{shundyak_force_2007} for frictional packings. To go beyond mean-field, one can perform a rigid cluster decomposition and identify the location of the rigidity transition: this occurs when at least one  rigid cluster spans the system. For frictionless packings, Laman's theorem \cite{laman_graphs_1970} can be implemented using what is known as a $(2,3)$ pebble game \cite{jacobs_algorithm_1997}, where $2$ represents the number of degrees of freedom per particle (translational) and $3$ represents the number of trivial global degrees of freedom. To do so, the packings are mapped to a constraint network where nodes (particle centers) are connected by a bond if their associated particles are in contact \cite{ellenbroek_rigidity_2015, ellenbroek_non-affine_2009, feng_percolation_1985, henkes_rigid_2016}. 

 In frictional packings ($\mu>0$), each particle has 2 translational and one rotational degrees of freedom.  The contacts between particles provide two constraints---one for the central (normal) force and one for the tangential force, except for contacts at the Coulomb threshold where normal and tangential forces are coupled, leading to only one independent constraint. To take that into account, the $(2,3)$ pebble game was recently extended \cite{henkes_rigid_2016} to a $(3,3)$ pebble game. It accommodates the additional rotational degree of freedom and the constraint network contains both single (sliding) bonds or double (frictional) bonds.

From the gluing approach used to build our sample, one concludes that playing a $(3,3)$ pebble game on a constraint network with all double bonds ($\mu = \infty$) maps out the rigid clusters. However, in frictional granular packings, local stresses will eventually be accommodated by contact slipping, and ultimately rearrangements. To account for accommodation in the stiff lattices, we make the following assumption based on experimental observations. We distinguish between contacts in the disordered lattice forming linear chains (only two contacts connect at each nodes),  and those that intersect at nodes with multiple other contacts. Since transverse forces can easily bend linear chains (unlike bonds having each node with degree $>2$, i.e. short links), we map each chain to a unique single bond. With this approximation, the constraint network has now changed from all double bonds to both single and double bonds, depending on the local coordination number.

Finally, to match the experimental boundary conditions, we introduce one particle representing a confining wall at the top and bottom of the system and connect them with single constraint bonds to the corresponding boundary nodes.

\subsection{Rigidity transition determines ductile vs. brittle failure}

Fig.~\ref{fig:clusters} shows sample results from applying the described pebble game algorithm on three intact disordered lattices. Each double (single) bond is indicated by a thick (thin) line. All identified rigid clusters are marked as colored subregions. For samples with $\Z < 3$, the analysis reveals predominantly-floppy networks, with only a few small isolated rigid clusters (see Fig.~\ref{fig:clusters}(a)). For samples with  $\Z>3$,  a rigid cluster percolates to the boundaries.  

\begin{figure}
\center
\includegraphics[width=0.7\columnwidth]{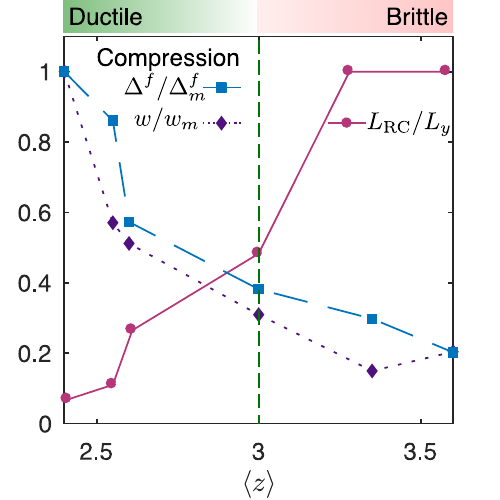}
\caption{{\bf Brittle-ductile vs. rigidity percolation transition.}
For compression experiments: vertical size of the largest rigid cluster obtained ($L_{\mathrm{RC}})$, normalized by the vertical system size ($L_y$), shown by magenta circles; average crack width $w$,  normalized by the maximum value ($w_m$), shown by purple diamonds; and loading force $\Delta^f$ at failure, normalized by its maximum value ($\Delta^f_m$), show by blue squares. The location of the isostatic condition $\Z_{\mathrm{iso}}= 3$, in infinite friction granular system, is marked by a vertical dashed line for comparison. }
\label{fig:transition}
\end{figure}

To highlight the connection between rigidity transition and the brittle-ductile transition, in Fig.~\ref{fig:transition} we  compare the $\Z$-evolution of the loading at failure ($\Delta^f$) and crack width ($w$), with the evolution of the vertical size of the largest rigid cluster ($L_{RC}$). Both $\Delta_f$ and $w$ decrease with $\Z$. For connectivity values $\Z>3$, a brittle behavior is reached. The largest cluster size evolves in the opposite manner: it increases until reaching the system size when $\Z>3$.
Therefore, samples for which a clear ductile failure behavior is identified are the ones for which there is a percolating floppy region with non-spanning rigid clusters. In contrast, brittle-like failure occurs at large $\Z$, in systems for which there is a percolating rigid cluster.

Finally, the geometry with $\Z = 3.0$ is of particular interest as it presents a floppy matrix but also large embedded rigid clusters. For this structure, the failure pattern was shown to exhibit narrow, crack-like, brittle features, while the failure process and characteristic loadings were more like the ductile samples. This mix of behaviors is echoed by the pebble game rigidity analysis below. Therefore, from the failure observation and rigidity analysis, the transition point is identified as close to $\Z\approx 3$, the infinite friction isostatic point
$\Z_{\mathrm{iso}}$.

\subsection{Rigidity signatures in the failure locations} 

For samples for which mesoscale rigid regions are identified  ($\Z = 3.0-3.15$), the rigid cluster decomposition additionally provides insights into the locations at which beams fail. For these networks, we isolate floppy vs. rigid regions. Since each beam failure corresponds to creating a new network by removing a bond from the previous, less damaged network, we re-apply the pebble game again after each failure event. Since the pebble game is highly non-local, a failure on one part of the lattice can cause a rigid edge in another part of the lattice to become floppy. Nonetheless, only a few edges evolve from rigid to floppy so that we here show the rigid clusters determined on the intact network.

\begin{figure}
\center
\includegraphics[width=0.99\columnwidth]{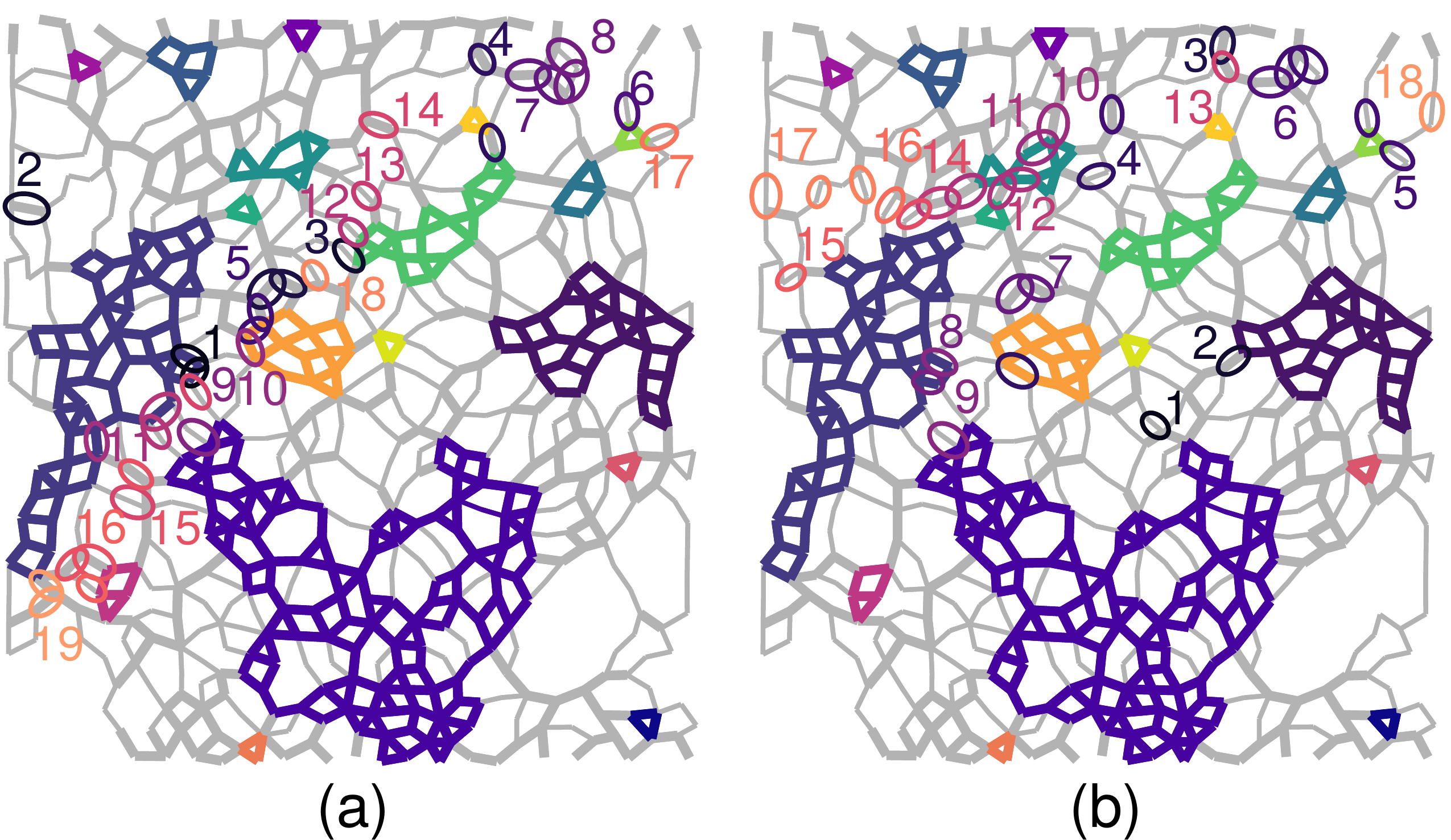}
\caption{{\bf Rigidity signatures in failure locations.}  Rigid cluster decomposition (colored networks) on the intact networks geometry
with  $\Z = 3.0$,  and failure locations of samples tested under (a) compression and (b) tension. Thick (thin) lines indicate double (single) bonds in the constraint network Failed edges are highlighted by ellipses, color-matched to the avalanche number indicative of the temporal ordering of failure events.
} 
\label{fig:locations}
\end{figure}

The rigid cluster decompositions are shown on Fig.~\ref{fig:locations} for the intact structure of degree $\Z = 3.0$ tested in (a) compression and (b) tension. The failure locations, which vary between samples and loadings directions, are numbered by their temporal ordering and indicated by ellipses. For the samples shown in Fig.~\ref{fig:locations} one (in panel $a$) and two (in panel $b$) failed bonds were initially rigid, but became floppy after a failure event elsewhere. 
The slight differences in geometry  result from beams that were broken during manufacturing, but these minor variations do not affect the rigid cluster decompositions which remain identical. 

\begin{table}
\begin{center}
\begin{tabular}{c c | c c}
\hline
~    & ~       & Floppy Failed  & System Floppy\\ 
$\Z$ & Loading & Edges  ($\%$)    &  Edges ($\%$)\\ \hline
3.0 & Compression & 83 & 57 \\ 
3.0 & Tension & 90 & 57 \\ 
3.0 & Tension & 100 & 53 \\
3.10 & Tension & 73 & 45 \\ 
3.15 & Tension & 74 & 44 \\ 
\end{tabular}
\caption{Forecasting efficiency of the rigid cluster decomposition for different samples. The first column indicates proportion of failed edges that are floppy, the second one lists the proportion of floppy edges in the initial network.}
\label{Table1}
\end{center}
\end{table}

Floppy edges represent about half of the total edges of the system for the samples with $\Z=3-3.15$. Failure events predominantly occur ($73-100\%$ of them) on the floppy edges separating rigid clusters. A physical explanation would be that floppy edges are freer to bend further, and thereby fail.  
In Table \ref{Table1}, we tabulate the proportion of failed edges that were marked as floppy for each sample, and the proportion of floppy edges of each intact system. 
These results indicate that in samples where both rigid and floppy clusters are present, the failure locations are not randomly distributed in the system. Instead, fractures seem to principally occur within the floppy regions determined by a rigidity analysis. Interestingly, because failures are occurring at locations outside of the rigid clusters, the clusters themselves do not evolve significantly as the damage progresses. 
On the other hand, because networks with connectivity  $\Z\gg 3$ have homogeneous-like structures that consist of a large rigid cluster, no information can be extracted for the failure locations. Indeed, breakages almost always occur on rigid bonds and a detailed analysis would be required to further understand the failure behavior of these networks. Similarly, for $\Z\ll3$, there is a large floppy matrix, and breakages occur primarily on floppy bonds.

\section*{Discussion and Conclusions} 

To study the failure locations and interpret the brittle-ductile transition, we have here considered a constraint-counting game that is rooted in how the samples were obtained. Yet, other games could be envisaged, in particular if one decides to focus on capturing the physical properties of the samples and assume that the network does not retain any information  about  the  particle  packing  from  which  it  was  derived.  In that case, in assigning the degrees of freedom and building the constraints network, one must account that the lattices studied here consist of \textit{welded} beams, which can be mechanically described as a spring network composed of two-body springs and three-body, or angular, springs between any two beams. The latter have been shown to stabilize networks below the central-force isostatic point \cite{feng_percolation_1984,feng_percolation_1984-1,kantor_elastic_1984,das_redundancy_2012}. In fact, an effective medium theory in Das \textit{et al.} \cite{das_redundancy_2012} has shown that the onset of rigidity percolation for welded beams correlates with the onset of connectivity percolation, which was initially argued in Kantor and Webman \cite{kantor_elastic_1984}. 

In this paper, we have shown that low- and high-connectivity networks differ dramatically in their response to uniaxial loading. Under both tension and compression loadings, a ductile-like response is present in low-connectivity samples, and a more brittle response is observed in high-connectivity samples. This behavior is consistently manifested in a number of quantifiable ways, from maximum load (larger for higher connectivity) to the width of cracks (diffuse for low-connectivity, narrow and localized for high connectivity).

Furthermore, we find that an adaptation of the frictional pebble game provides quantitative insight into how these two classes of disordered lattices fail.  In particular, low- and high-connectivity networks both have rigid and floppy sub-regions and the size of the rigid clusters correlates with the observation of the ductile vs. brittle behavior. Low-connectivity networks are dominated by floppy regions in which many failure locations are possible, while high-connectivity networks are dominated by rigid clusters where one failure can trigger a cascade of nearby failures within rigid clusters. Intriguingly, we additionally observe that in samples with both rigid and (sometimes newly-created) floppy regions, the edge failures are more likely to occur along the floppy regions separating rigid clusters. 

This observation can be tested in numerical simulations for a variety of networks.  A systematic study would permit testing its validity for structures of geometry constructed from different origins, i.e. not only from visible force-chains in frictional packings. Identifying the possible failure locations also raises the possibility of designing disordered structures with pre-determined failure behaviors. Finally, a possible connection between the failure locations observed in the lattices and contact rearrangements or force-chain buckling in the granular packings remains to be investigated. 

\textbf{Acknowledgements:} We are grateful to support from the James S. McDonnell Foundation, National Science Foundation Grant DMR-1206808, and DMR-1507938.

\textbf{Author contributions:}
J.E.K. performed the granular experiments and extracted the force chain networks;
E.B. created the lattices and performed, processed, and analyzed the experiments;
S.E.H, J.M.S. wrote the pebble-game code;
K.L., E.B. performed the pebble-game analysis;
E.B., J.M.S, S.E.H, and K.D. connected the theoretical and experimental results;
E.B., J.M.S, S.E.H, and K.D. wrote the manuscript. 

\textbf{Competing interests:}. The authors have no competing interests to declare. 

\textbf{Materials \& Correspondence:} Estelle Berthier (ehberthi@ncsu.edu)


\begin{thebibliography}{10}
\expandafter\ifx\csname url\endcsname\relax
  \def\url#1{\texttt{#1}}\fi
\expandafter\ifx\csname urlprefix\endcsname\relax\def\urlprefix{URL }\fi
\providecommand{\bibinfo}[2]{#2}
\providecommand{\eprint}[2][]{\url{#2}}

\bibitem{oshima_extraordinary_2018}
\bibinfo{author}{Oshima, Y.}, \bibinfo{author}{Nakamura, A.} \&
  \bibinfo{author}{Matsunaga, K.}
\newblock \bibinfo{title}{Extraordinary plasticity of an inorganic
  semiconductor in darkness}.
\newblock \emph{\bibinfo{journal}{Science}} \textbf{\bibinfo{volume}{360}},
  \bibinfo{pages}{772--774} (\bibinfo{year}{2018}).

\bibitem{arif_spontaneous_2012}
\bibinfo{author}{Arif, S.}, \bibinfo{author}{Tsai, J.-C.} \&
  \bibinfo{author}{Hilgenfeldt, S.}
\newblock \bibinfo{title}{Spontaneous brittle-to-ductile transition in aqueous
  foam}.
\newblock \emph{\bibinfo{journal}{Journal of Rheology}}
  \textbf{\bibinfo{volume}{56}}, \bibinfo{pages}{485--499}
  (\bibinfo{year}{2012}).

\bibitem{jang_ductile-brittle_1984}
\bibinfo{author}{Jang, B.~Z.}, \bibinfo{author}{Uhlmann, D.~R.} \&
  \bibinfo{author}{Vander~Sande, J.~B.}
\newblock \bibinfo{title}{Ductile-brittle transition in polymers.}
\newblock \emph{\bibinfo{journal}{Journal of Applied Polymer Science}}
  \textbf{\bibinfo{volume}{29}}, \bibinfo{pages}{3409--3420}
  (\bibinfo{year}{1984}).

\bibitem{Tanguy2005}
\bibinfo{author}{Tanguy, B.}, \bibinfo{author}{Besson, J.},
  \bibinfo{author}{Piques, R.} \& \bibinfo{author}{Pineau, A.}
\newblock \bibinfo{title}{Ductile to brittle transition of an a508 steel
  characterized by charpy impact test: Part {I}: experimental results}.
\newblock \emph{\bibinfo{journal}{Engineering Fracture Mechanics}}
  \textbf{\bibinfo{volume}{72}}, \bibinfo{pages}{49 -- 72}
  (\bibinfo{year}{2005}).

\bibitem{wong_brittle-ductile_2012}
\bibinfo{author}{Wong, T.-F.} \& \bibinfo{author}{Baud, P.}
\newblock \bibinfo{title}{The brittle-ductile transition in porous rock: {A}
  review}.
\newblock \emph{\bibinfo{journal}{Journal of Structural Geology}}
  \textbf{\bibinfo{volume}{44}}, \bibinfo{pages}{25--53}
  (\bibinfo{year}{2012}).

\bibitem{roux_rupture_1988}
\bibinfo{author}{Roux, S.}, \bibinfo{author}{Hansen, A.},
  \bibinfo{author}{Herrmann, H.} \& \bibinfo{author}{Guyon, E.}
\newblock \bibinfo{title}{Rupture of heterogeneous media in the limit of
  infinite disorder}.
\newblock \emph{\bibinfo{journal}{Journal of Statistical Physics}}
  \textbf{\bibinfo{volume}{52}}, \bibinfo{pages}{237--244}
  (\bibinfo{year}{1988}).

\bibitem{alava_role_2008}
\bibinfo{author}{Alava, M.~J.}, \bibinfo{author}{Nukala, P. K. V.~V.} \&
  \bibinfo{author}{Zapperi, S.}
\newblock \bibinfo{title}{Role of disorder in the size scaling of material
  strength}.
\newblock \emph{\bibinfo{journal}{Physical Review Letters}}
  \textbf{\bibinfo{volume}{100}}, \bibinfo{pages}{055502}
  (\bibinfo{year}{2008}).

\bibitem{shekhawat_damage_2013}
\bibinfo{author}{Shekhawat, A.}, \bibinfo{author}{Zapperi, S.} \&
  \bibinfo{author}{Sethna, J.~P.}
\newblock \bibinfo{title}{From damage percolation to crack nucleation through
  finite size criticality}.
\newblock \emph{\bibinfo{journal}{Physical Review Letters}}
  \textbf{\bibinfo{volume}{110}}, \bibinfo{pages}{185505}
  (\bibinfo{year}{2013}).

\bibitem{curtin_brittle_1990}
\bibinfo{author}{Curtin, W.~A.} \& \bibinfo{author}{Scher, H.}
\newblock \bibinfo{title}{Brittle fracture in disordered materials: A spring
  network model}.
\newblock \emph{\bibinfo{journal}{Journal of Materials Research}}
  \textbf{\bibinfo{volume}{5}}, \bibinfo{pages}{535--553}
  (\bibinfo{year}{1990}).

\bibitem{kahng_electrical_1988}
\bibinfo{author}{Kahng, B.}, \bibinfo{author}{Batrouni, G.~G.},
  \bibinfo{author}{Redner, S.}, \bibinfo{author}{de~Arcangelis, L.} \&
  \bibinfo{author}{Herrmann, H.~J.}
\newblock \bibinfo{title}{Electrical breakdown in a fuse network with random,
  continuously distributed breaking strengths}.
\newblock \emph{\bibinfo{journal}{Phys. Rev. B}} \textbf{\bibinfo{volume}{37}},
  \bibinfo{pages}{7625--7637} (\bibinfo{year}{1988}).

\bibitem{hanifpour_mechanics_2018}
\bibinfo{author}{Hanifpour, M.}, \bibinfo{author}{Petersen, C.~F.},
  \bibinfo{author}{Alava, M.~J.} \& \bibinfo{author}{Zapperi, S.}
\newblock \bibinfo{title}{Mechanics of disordered auxetic metamaterials}.
\newblock \emph{\bibinfo{journal}{The European Physical Journal B}}
  \textbf{\bibinfo{volume}{91}}, \bibinfo{pages}{271} (\bibinfo{year}{2018}).

\bibitem{driscoll_role_2016}
\bibinfo{author}{Driscoll, M.~M.} \emph{et~al.}
\newblock \bibinfo{title}{The role of rigidity in controlling material
  failure}.
\newblock \emph{\bibinfo{journal}{Proceedings of the National Academy of
  Sciences}} \textbf{\bibinfo{volume}{113}}, \bibinfo{pages}{10813--10817}
  (\bibinfo{year}{2016}).

\bibitem{zhang_fiber_2017}
\bibinfo{author}{Zhang, L.}, \bibinfo{author}{Rocklin, D.~Z.},
  \bibinfo{author}{Sander, L.~M.} \& \bibinfo{author}{Mao, X.}
\newblock \bibinfo{title}{Fiber networks below the isostatic point: Fracture
  without stress concentration}.
\newblock \emph{\bibinfo{journal}{Phys. Rev. Materials}}
  \textbf{\bibinfo{volume}{1}}, \bibinfo{pages}{052602} (\bibinfo{year}{2017}).

\bibitem{bouzid_network_2017}
\bibinfo{author}{Bouzid, M.} \& \bibinfo{author}{Del~Gado, E.}
\newblock \bibinfo{title}{Network topology in soft gels: Hardening and
  softening materials}.
\newblock \emph{\bibinfo{journal}{Langmuir}} \textbf{\bibinfo{volume}{34}},
  \bibinfo{pages}{773--781} (\bibinfo{year}{2018}).

\bibitem{Paulose7639}
\bibinfo{author}{Paulose, J.}, \bibinfo{author}{Meeussen, A.~S.} \&
  \bibinfo{author}{Vitelli, V.}
\newblock \bibinfo{title}{Selective buckling via states of self-stress in
  topological metamaterials}.
\newblock \emph{\bibinfo{journal}{Proceedings of the National Academy of
  Sciences}} \textbf{\bibinfo{volume}{112}}, \bibinfo{pages}{7639--7644}
  (\bibinfo{year}{2015}).

\bibitem{bertoldi_flexible_2017}
\bibinfo{author}{Bertoldi, K.}, \bibinfo{author}{Vitelli, V.},
  \bibinfo{author}{Christensen, J.} \& \bibinfo{author}{van Hecke, M.}
\newblock \bibinfo{title}{Flexible mechanical metamaterials}.
\newblock \emph{\bibinfo{journal}{Nature Reviews Materials}}
  \textbf{\bibinfo{volume}{2}}, \bibinfo{pages}{17066} (\bibinfo{year}{2017}).

\bibitem{goodrich_principle_2015}
\bibinfo{author}{Goodrich, C.~P.}, \bibinfo{author}{Liu, A.~J.} \&
  \bibinfo{author}{Nagel, S.~R.}
\newblock \bibinfo{title}{The principle of independent bond-level response:
  tuning by pruning to exploit disorder for global behavior}.
\newblock \emph{\bibinfo{journal}{Physical Review Letters}}
  \textbf{\bibinfo{volume}{114}}, \bibinfo{pages}{225501}
  (\bibinfo{year}{2015}).

\bibitem{reid_auxetic_2018}
\bibinfo{author}{Reid, D.~R.} \emph{et~al.}
\newblock \bibinfo{title}{Auxetic metamaterials from disordered networks}.
\newblock \emph{\bibinfo{journal}{Proceedings of the National Academy of
  Sciences}} \textbf{\bibinfo{volume}{115}}, \bibinfo{pages}{E1384--E1390}
  (\bibinfo{year}{2018}).

\bibitem{rocks_designing_2017}
\bibinfo{author}{Rocks, J.~W.} \emph{et~al.}
\newblock \bibinfo{title}{Designing allostery-inspired response in mechanical
  networks}.
\newblock \emph{\bibinfo{journal}{Proceedings of the National Academy of
  Sciences}} \textbf{\bibinfo{volume}{114}}, \bibinfo{pages}{2520--2525}
  (\bibinfo{year}{2017}).

\bibitem{alava_statistical_2006}
\bibinfo{author}{Alava, M.~J.}, \bibinfo{author}{Nukala, P. K. V.~V.} \&
  \bibinfo{author}{Zapperi, S.}
\newblock \bibinfo{title}{Statistical models of fracture}.
\newblock \emph{\bibinfo{journal}{Advances in Physics}}
  \textbf{\bibinfo{volume}{55}}, \bibinfo{pages}{349--476}
  (\bibinfo{year}{2006}).

\bibitem{de_arcangelis_random_1985}
\bibinfo{author}{de~Arcangelis, L.}, \bibinfo{author}{Redner, S.} \&
  \bibinfo{author}{Herrmann, H.}
\newblock \bibinfo{title}{A random fuse model for breaking processes}.
\newblock \emph{\bibinfo{journal}{Journal de Physique Lettres}}
  \textbf{\bibinfo{volume}{46}}, \bibinfo{pages}{585--590}
  (\bibinfo{year}{1985}).

\bibitem{gilabert_random_1987}
\bibinfo{author}{Gilabert, A.}, \bibinfo{author}{Vanneste, C.},
  \bibinfo{author}{Sornette, D.} \& \bibinfo{author}{Guyon, E.}
\newblock \bibinfo{title}{The random fuse network as a model of rupture in a
  disordered medium}.
\newblock \emph{\bibinfo{journal}{Journal de Physique}}
  \textbf{\bibinfo{volume}{48}}, \bibinfo{pages}{763--770}
  (\bibinfo{year}{1987}).

\bibitem{pradhan_failure_2010}
\bibinfo{author}{Pradhan, S.}, \bibinfo{author}{Hansen, A.} \&
  \bibinfo{author}{Chakrabarti, B.~K.}
\newblock \bibinfo{title}{Failure processes in elastic fiber bundles}.
\newblock \emph{\bibinfo{journal}{Reviews of Modern Physics}}
  \textbf{\bibinfo{volume}{82}}, \bibinfo{pages}{499--555}
  (\bibinfo{year}{2010}).

\bibitem{Kun2006}
\bibinfo{author}{Kun, F.}, \bibinfo{author}{Raischel, F.},
  \bibinfo{author}{Hidalgo, R.} \& \bibinfo{author}{Herrmann, H.}
\newblock \emph{\bibinfo{title}{Extensions of Fibre Bundle Models}},
  \bibinfo{pages}{57--92} (\bibinfo{publisher}{Springer Berlin Heidelberg},
  \bibinfo{address}{Berlin, Heidelberg}, \bibinfo{year}{2006}).

\bibitem{nukala_percolation_2004}
\bibinfo{author}{Nukala, P. K. V.~V.}, \bibinfo{author}{{\v S}imunovi{\' c},
  S.} \& \bibinfo{author}{Zapperi, S.}
\newblock \bibinfo{title}{Percolation and localization in the random fuse
  model}.
\newblock \emph{\bibinfo{journal}{Journal of Statistical Mechanics: Theory and
  Experiment}} \textbf{\bibinfo{volume}{2004}}, \bibinfo{pages}{P08001}
  (\bibinfo{year}{2004}).

\bibitem{ray_breakdown_2006}
\bibinfo{author}{Ray, P.}
\newblock \bibinfo{title}{Breakdown of heterogeneous materials}.
\newblock \emph{\bibinfo{journal}{Computational Materials Science}}
  \textbf{\bibinfo{volume}{37}}, \bibinfo{pages}{141--145}
  (\bibinfo{year}{2006}).

\bibitem{jekollmer_pegs:_2018}
\bibinfo{author}{Kollmer, J.~E.}
\newblock \bibinfo{title}{{PEGS}: {Photo}-elastic {Grain} {Solver}}
  (\bibinfo{year}{2018}).
\newblock \urlprefix\url{https://github.com/jekollmer/PEGS}.

\bibitem{daniels_photoelastic_2017}
\bibinfo{author}{Daniels, K.~E.}, \bibinfo{author}{Kollmer, J.~E.} \&
  \bibinfo{author}{Puckett, J.~G.}
\newblock \bibinfo{title}{Photoelastic force measurements in granular
  materials}.
\newblock \emph{\bibinfo{journal}{Review of Scientific Instruments}}
  \textbf{\bibinfo{volume}{88}}, \bibinfo{pages}{051808}
  (\bibinfo{year}{2017}).

\bibitem{broedersz_criticality_2011}
\bibinfo{author}{Broedersz, C.~P.}, \bibinfo{author}{Mao, X.},
  \bibinfo{author}{Lubensky, T.~C.} \& \bibinfo{author}{MacKintosh, F.~C.}
\newblock \bibinfo{title}{Criticality and isostaticity in fibre networks}.
\newblock \emph{\bibinfo{journal}{Nature Physics}}
  \textbf{\bibinfo{volume}{7}}, \bibinfo{pages}{983--988}
  (\bibinfo{year}{2011}).

\bibitem{ellenbroek_rigidity_2015}
\bibinfo{author}{Ellenbroek, W.~G.}, \bibinfo{author}{Hagh, V.~F.},
  \bibinfo{author}{Kumar, A.}, \bibinfo{author}{Thorpe, M.~F.} \&
  \bibinfo{author}{van Hecke, M.}
\newblock \bibinfo{title}{Rigidity loss in disordered systems: three
  scenarios}.
\newblock \emph{\bibinfo{journal}{Physical Review Letters}}
  \textbf{\bibinfo{volume}{114}}, \bibinfo{pages}{135501}
  (\bibinfo{year}{2015}).

\bibitem{feng_percolation_1984}
\bibinfo{author}{Feng, S.} \& \bibinfo{author}{Sen, P.~N.}
\newblock \bibinfo{title}{Percolation on elastic networks - new exponent and
  threshold}.
\newblock \emph{\bibinfo{journal}{Physical Review Letters}}
  \textbf{\bibinfo{volume}{52}}, \bibinfo{pages}{216--219}
  (\bibinfo{year}{1984}).

\bibitem{ellenbroek_non-affine_2009}
\bibinfo{author}{Ellenbroek, W.~G.}, \bibinfo{author}{Zeravcic, Z.},
  \bibinfo{author}{van Saarloos, W.} \& \bibinfo{author}{van Hecke, M.}
\newblock \bibinfo{title}{Non-affine response: jammed packings vs. spring
  networks}.
\newblock \emph{\bibinfo{journal}{EPL (Europhysics Letters)}}
  \textbf{\bibinfo{volume}{87}}, \bibinfo{pages}{34004} (\bibinfo{year}{2009}).

\bibitem{liu_jamming_2010}
\bibinfo{author}{Liu, A.~J.} \& \bibinfo{author}{Nagel, S.~R.}
\newblock \bibinfo{title}{The jamming transition and the marginally jammed
  solid}.
\newblock \emph{\bibinfo{journal}{Annual Review of Condensed Matter Physics}}
  \textbf{\bibinfo{volume}{1}}, \bibinfo{pages}{347--369}
  (\bibinfo{year}{2010}).

\bibitem{van_hecke_jamming_2010}
\bibinfo{author}{van Hecke, M.}
\newblock \bibinfo{title}{Jamming of soft particles: geometry, mechanics,
  scaling and isostaticity}.
\newblock \emph{\bibinfo{journal}{Journal of Physics: Condensed Matter}}
  \textbf{\bibinfo{volume}{22}}, \bibinfo{pages}{033101}
  (\bibinfo{year}{2010}).

\bibitem{majmudar_contact_2005}
\bibinfo{author}{Majmudar, T.~S.} \& \bibinfo{author}{Behringer, R.~P.}
\newblock \bibinfo{title}{Contact force measurements and stress-induced
  anisotropy in granular materials}.
\newblock \emph{\bibinfo{journal}{Nature}} \textbf{\bibinfo{volume}{435}},
  \bibinfo{pages}{1079--1082} (\bibinfo{year}{2005}).

\bibitem{puckett_equilibrating_2013}
\bibinfo{author}{Puckett, J.~G.} \& \bibinfo{author}{Daniels, K.~E.}
\newblock \bibinfo{title}{Equilibrating temperature like variables in jammed
  granular subsystems}.
\newblock \emph{\bibinfo{journal}{Phys. Rev. Lett.}}
  \textbf{\bibinfo{volume}{110}}, \bibinfo{pages}{058001}
  (\bibinfo{year}{2013}).

\bibitem{kollmer_betweenness_2018}
\bibinfo{author}{Kollmer, J.~E.} \& \bibinfo{author}{Daniels, K.~E.}
\newblock \bibinfo{title}{Betweenness centrality as predictor for forces in
  granular packings}.
\newblock \emph{\bibinfo{journal}{{Preprint} at
  https://arxiv.org/abs/1807.01786}}  (\bibinfo{year}{2018}).

\bibitem{bililign2018}
\bibinfo{author}{Bililign, E.~S.}, \bibinfo{author}{Kollmer, J.~E.} \&
  \bibinfo{author}{Daniels, K.~E.}
\newblock \bibinfo{title}{Protocol-dependence and state variables in the
  force-moment ensemble}.
\newblock \emph{\bibinfo{journal}{({Physical} {Review} {Letters}, in press)
  {Preprint} at https://arxiv.org/abs/1802.09641}}  (\bibinfo{year}{2018}).

\bibitem{moukarzel_stressed_1995}
\bibinfo{author}{Moukarzel, C.} \& \bibinfo{author}{Duxbury, P.~M.}
\newblock \bibinfo{title}{Stressed backbone and elasticity of random
  central-force systems}.
\newblock \emph{\bibinfo{journal}{Physical Review Letters}}
  \textbf{\bibinfo{volume}{75}}, \bibinfo{pages}{4055--4058}
  (\bibinfo{year}{1995}).

\bibitem{jacobs_generic_1995}
\bibinfo{author}{Jacobs, D.~J.} \& \bibinfo{author}{Thorpe, M.~F.}
\newblock \bibinfo{title}{Generic rigidity percolation: the pebble game}.
\newblock \emph{\bibinfo{journal}{Physical Review Letters}}
  \textbf{\bibinfo{volume}{75}}, \bibinfo{pages}{4051--4054}
  (\bibinfo{year}{1995}).

\bibitem{feng_percolation_1985}
\bibinfo{author}{Feng, S.}
\newblock \bibinfo{title}{Percolation properties of granular elastic networks
  in two dimensions}.
\newblock \emph{\bibinfo{journal}{Physical Review B}}
  \textbf{\bibinfo{volume}{32}}, \bibinfo{pages}{510--513}
  (\bibinfo{year}{1985}).

\bibitem{henkes_rigid_2016}
\bibinfo{author}{Henkes, S.}, \bibinfo{author}{Quint, D.~A.},
  \bibinfo{author}{Fily, Y.} \& \bibinfo{author}{Schwarz, J.}
\newblock \bibinfo{title}{Rigid cluster decomposition reveals criticality in
  frictional jamming}.
\newblock \emph{\bibinfo{journal}{Physical Review Letters}}
  \textbf{\bibinfo{volume}{116}}, \bibinfo{pages}{028301}
  (\bibinfo{year}{2016}).

\bibitem{jacobs_algorithm_1997}
\bibinfo{author}{Jacobs, D.~J.} \& \bibinfo{author}{Hendrickson, B.}
\newblock \bibinfo{title}{An algorithm for two-dimensional rigidity
  percolation: the pebble game}.
\newblock \emph{\bibinfo{journal}{Journal of Computational Physics}}
  \textbf{\bibinfo{volume}{137}}, \bibinfo{pages}{346--365}
  (\bibinfo{year}{1997}).

\bibitem{maxwell_l._1864}
\bibinfo{author}{Maxwell, J.~C.}
\newblock \bibinfo{title}{On the calculation of the equilibrium and stiffness
  of frames}.
\newblock \emph{\bibinfo{journal}{The London, Edinburgh, and Dublin
  Philosophical Magazine and Journal of Science}}
  \textbf{\bibinfo{volume}{27}}, \bibinfo{pages}{294--299}
  (\bibinfo{year}{1864}).

\bibitem{shundyak_force_2007}
\bibinfo{author}{Shundyak, K.}, \bibinfo{author}{van Hecke, M.} \&
  \bibinfo{author}{van Saarloos, W.}
\newblock \bibinfo{title}{Force mobilization and generalized isostaticity in
  jammed packings of frictional grains}.
\newblock \emph{\bibinfo{journal}{Phys. Rev. E}} \textbf{\bibinfo{volume}{75}},
  \bibinfo{pages}{010301} (\bibinfo{year}{2007}).

\bibitem{laman_graphs_1970}
\bibinfo{author}{Laman, G.}
\newblock \bibinfo{title}{On graphs and rigidity of plane skeletal structures}.
\newblock \emph{\bibinfo{journal}{Journal of Engineering Mathematics}}
  \textbf{\bibinfo{volume}{4}}, \bibinfo{pages}{331--340}
  (\bibinfo{year}{1970}).

\bibitem{feng_percolation_1984-1}
\bibinfo{author}{Feng, S.}, \bibinfo{author}{Sen, P.~N.},
  \bibinfo{author}{Halperin, B.~I.} \& \bibinfo{author}{Lobb, C.~J.}
\newblock \bibinfo{title}{Percolation on two-dimensional elastic networks with
  rotationally invariant bond-bending forces}.
\newblock \emph{\bibinfo{journal}{Physical Review B}}
  \textbf{\bibinfo{volume}{30}}, \bibinfo{pages}{5386--5389}
  (\bibinfo{year}{1984}).

\bibitem{kantor_elastic_1984}
\bibinfo{author}{Kantor, Y.} \& \bibinfo{author}{Webman, I.}
\newblock \bibinfo{title}{Elastic {Properties} of {Random} {Percolating}
  {Systems}}.
\newblock \emph{\bibinfo{journal}{Physical Review Letters}}
  \textbf{\bibinfo{volume}{52}}, \bibinfo{pages}{1891--1894}
  (\bibinfo{year}{1984}).

\bibitem{das_redundancy_2012}
\bibinfo{author}{Das, M.}, \bibinfo{author}{Quint, D.~A.} \&
  \bibinfo{author}{Schwarz, J.~M.}
\newblock \bibinfo{title}{Redundancy and cooperativity in the mechanics of
  compositely crosslinked filamentous networks}.
\newblock \emph{\bibinfo{journal}{PLoS ONE}} \textbf{\bibinfo{volume}{7}},
  \bibinfo{pages}{e35939} (\bibinfo{year}{2012}).

\end{thebibliography}
\end{document}